\documentclass[twocolumn,twoside,slac_two]{revtex4}
\usepackage{graphicx}
\usepackage{fancyhdr}
\usepackage{amsmath}
\usepackage{cancel}
\def\kslash {k\kern-.5em\slash}
\begin{document}

\def\pslash {p\kern-.5em\slash}
\def \kslash {k\kern-.5em\slash}
\def\qslash {q\kern-.5em\slash}
\def\eslash {\epsilon\kern-.5em\slash}
\def\Dslash {D\kern-.7em\slash}
\def\dslash {\partial\kern-.7em\slash}
\def\idot{\!\cdot\!}
\def\kv{\mathbf{k}}
\def\uv{\mathbf{u}}
\def\pv{\mathbf{p}}
\def\qv{\mathbf{q}}
\def\ra{{\rm I}}
\def\rb{{\rm I\!I}}
\def\rc{{\rm I\!I\!I}}
\def\rd{{\rm I\!V}}
\def\rda{{{\rm I\!V}_{\!1}}}
\def\rdb{{{\rm I\!V}_{\!2}}}
\def\pra{\phantom{\rm I}}
\def\prb{\phantom{\rm I\!I}}
\def\prc{\phantom{\rm I\!I\!I}}
\def\prd{\phantom{\rm I\!V}}
\def\prda{{\phantom{{\rm I\!V}_{\!1}}}}
\def\prdb{{\phantom{{\rm I\!V}_{\!2}}}}
\def\pip{p_{\!i\!\perp}}
\def\pap{p_{\!a\!\perp}}
\def\pbp{p_{\!b\!\perp}}
\def\c{,}
\def\gm{\gamma}
\def\gmf{\gamma_5}
\def\hs{\widehat{\Sigma}}
\def\fw{\,}
\def\bk{\!}
\def\.{ \! \cdot \! }
\def\ee{{{\bf e}_{0}}}
\def\ex{{{\bf e}_{1}}}
\def\ey{{{\bf e}_{2}}}
\def\ez{{{\bf e}_{3}}}
\def\A{\mathcal{A}}
\def\N{\mathbb{N}}
\def\M{\mathbb{M}}
\def\F{\mathcal{F}}
\def\P{\mathcal{P}}
\setlength{\unitlength}{1mm}
\title{Electroweak radiative corrections \\to neutrino--nucleon scattering at NuTeV}

%

\author{Kwangwoo Park}
\email{kpark@smu.edu}
\affiliation{Department of Physics, Southern Methodist University, Dallas, TX 75275, USA}
\author{Ulrich Baur}
\email{baur@ubhep.physics.buffalo.edu}
\affiliation{Department of Physics, SUNY at Buffalo, Buffalo, NY 14260, USA}
\author{Doreen Wackeroth}
\email{dow@ubpheno.physics.buffalo.edu}
\affiliation{Department of Physics, SUNY at Buffalo, Buffalo, NY 14260,
  USA \\ Institut f\"ur Theoretische Teilchenphysik, Karlsruhe Institute of Technology (KIT), Universit\"at Karlsruhe, D-76128 Karlsruhe, Germany}

\begin{abstract}
  A dedicated effort by both the experimental and theoretical communities is
  crucial for achieving a precise determination of Standard Model parameters
  such as the $W$ mass ($M_W$). $M_W$ is measured directly at the CERN LEP2
  $e^+ e^-$ and the Fermilab Tevatron $p \bar p$ colliders, resulting in a
  precision of $\delta M_W/M_W=0.03\%$~\cite{ewwg}.  A complementary $M_W$
  measurement is provided by the NuTeV
  collaboration~\cite{Zeller:2001hh,Zeller:2002dx}, which extract $\sin^2
  \theta_W$, and thus $M_W$, from the ratio of deep-inelastic neutral and
  charged-current neutrino(anti-neutrino)-Nucleon ($\nu N(\bar \nu N)$)
  scattering cross sections. However, their result differs from direct
  measurements performed at LEP2 and the Tevatron by about three standard
  deviations~\cite{Zeller:2001hh,Zeller:2002dx}.  Possible sources for the
  origin of this discrepancy have been extensively studied in the literature
  (see, e.~g.,~\cite{McFarland:2003jw}), among them the impact of electroweak
  radiative corrections~\cite{Diener:2003ss,Arbuzov:2004zr,Diener:2005me}.
  Here we provide first (preliminary) results of a new calculation of
  electroweak ${\cal O}(\alpha)$ corrections with emphasis on the effects of
  non-zero muon and charm quark masses.  We find non-negligible shifts in
  $\sin^2\theta_W$ due to these mass effects but more detailed studies
  including detector resolution effects are needed to determine their impact
  on $M_W$ as extracted by the NuTeV collaboration.
\end{abstract}

\maketitle

\thispagestyle{fancy}


\section{Introduction}

The Standard Model (SM) represents the best current understanding of
electroweak and strong interactions of elementary particles. In recent years
it has been impressively confirmed experimentally through the precise
determination of $W$ and $Z$ boson properties at the CERN LEP and the Stanford
Linear $e^+ e^-$ colliders, and the discovery of the top quark at the Fermilab
Tevatron $p \overline{p}$ collider.

\begin{figure}
\begin{center}
\includegraphics[width=0.4 \textwidth]{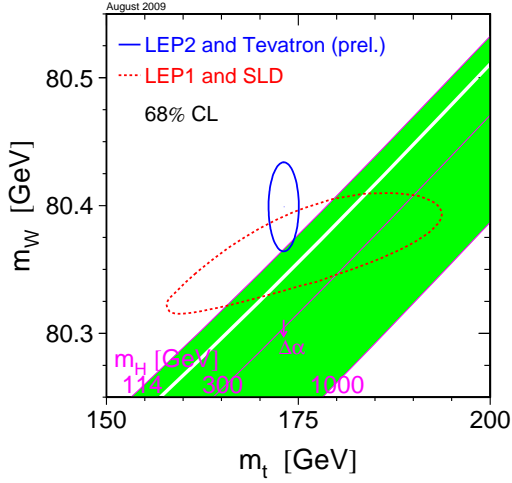}
\end{center}
\caption{The SM prediction for $M_W$ with dependence on the top--quark
  mass ($M_t$) and Higgs boson mass ($M_H$), resulting in the shaded
  band, is compared with the experimental values of $M_W$ and $M_t$
  (solid ellipse) and an indirect measurement from all electroweak
  precision data (dotted ellipse)~\cite{ewwg}. Present values of $M_W$,
  and $M_t$ favor a relatively light SM Higgs boson, while the NuTeV
  value of $M_W (=80.136\pm 0.084 \, {\rm GeV})$~\cite{Zeller:2001hh,Zeller:2002dx}
  prefers a much higher Higgs boson mass.}\label{MwMhMt}
\end{figure}

A precise measurement of $M_W$ does not only provide a further precisely known
SM input parameter, but significantly improves the indirect limit on the
Higgs-boson mass obtained by comparing SM predictions with electroweak
precision data as illustrated in Fig.~\ref{MwMhMt}

\begin{figure}
\begin{center}
\includegraphics[width=0.3 \textwidth]{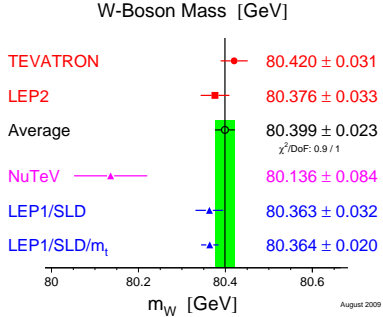}
\end{center}
\caption{Direct (Tevatron, LEP2, and NuTeV) and indirect measurements
  of $M_W$. The NuTeV value of $M_W$ differs from the world average
  value by about $3 \sigma$~\cite{ewwg}.}\label{Mwave}
\end{figure}

A measurement of $M_W$ can also be extracted from a measurement of the sine
squared of the weak mixing angle, $\sin^2 \theta_W$, via the well-known
relation between the $W$ and $Z$ boson masses, $M_W^2=M_Z^2 (1-\sin^2
\theta_W)$.  The NuTeV collaboration extracts $M_W$ from the ratio of neutral
and charged-current neutrino and anti-neutrino cross
sections~\cite{Zeller:2001hh,Zeller:2002dx}. Their results differ from direct
measurements performed at LEP2 and the Tevatron by about $3
\sigma$~\cite{Zeller:2001hh,Zeller:2002dx} as shown in Fig.~\ref{Mwave}.

Much effort, both experimentally and theoretically, has gone into
understanding this discrepancy. These efforts include studies of QCD
corrections, parton distribution functions~\cite{Martin:2004dh,Ball:2009mk},
and nuclear structure (see, e.~g.,~\cite{McFarland:2003jw} for an overview).
However, the impact of electroweak radiative corrections has not been fully
studied yet. In the extraction of $M_W$ from NuTeV data, only part of the
electroweak corrections have been included~\cite{Bardin:1986bc}. Since then
the complete calculation of these corrections has been made available in the
literature~\cite{Diener:2003ss,Arbuzov:2004zr,Diener:2005me}, but a realistic,
experimental study of their impact on the NuTeV measurement on $M_W$ has not
been performed yet.

In order to remedy this situation we calculated the
complete $\mathcal{O}(\alpha)$ contribution to neutrino--nucleon
scattering including the full muon and charm-quark mass dependence,
which has been neglected in previous studies. Here we present first
preliminary results of this new calculation with emphasis on the
above-mentioned 
mass effects. A detailed study, also taking into account more realistic
detector resolution effects, is in progress~\cite{paper}.

\section{Some details of the calculation}

\begin{figure}
\begin{center}
\includegraphics[width=0.4 \textwidth]{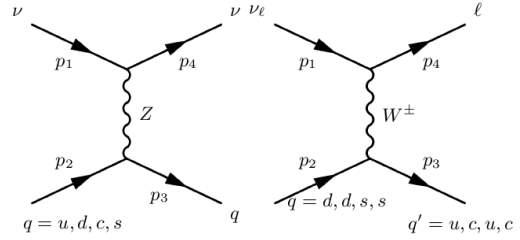}
\end{center}
\caption{Feynman diagrams for the tree-level NC (left) and CC (right) $\nu N$
  scattering processes.}
\label{tree}
\end{figure}

Our calculation of the complete ${\cal O}(\alpha)$ corrections to the
neutral-current (NC) and charged-current (CC) $\nu N(\bar \nu N)$ scattering
processes (the tree-level Feynman diagrams are shown in Fig.~\ref{tree})
follows closely the treatment of $s$-channel $W$ and $Z$ production at hadron
colliders of~\cite{Baur:1998kt,Baur:2001ze}. The ${\cal O}(\alpha)$ corrections consist of
the full set of electroweak one--loop diagrams and real photon radiation from
both the external charged fermion legs and the internal $W$ boson in the CC
process. As usual, they exhibit UV and IR divergences. UV divergences are
canceled by including the counterterms of the on-shell renormalization
scheme~\cite{Bohm:1986rj,Denner:1991kt}. By applying the two-cutoff
phase-space-slicing method~\cite{Harris:2001sx}, we extract the soft and
collinear singularities from the real photonic corrections. We use fermion
masses and a fictitious photon mass as regulators for the soft
and collinear singularities. The photon mass dependence cancels in the sum of
virtual and real soft-photon radiation, but mass singularities of the form
$\log(\hat t/m_f^2)$ may survive, which arise when the photon is emitted
collinear with the charged fermion. In the case of final-state photon
radiation, in inclusive observables these mass singularities cancel. However, mass singularities connected to
initial-state photon radiation survive in general. These need to be absorbed
in the parton distribution functions (PDF), which can be done in analogy to
gluon radiation in QCD.  Finally, the numerical phase space integration was
done using Monte Carlo integration techniques based on the Vegas
algorithm~\cite{Lepage:1977sw}.

After convolution with the quark PDFs, the predictions for the hadronic,
electroweak (EW) next-to-leading order (NLO) cross section for $\nu N$
scattering is obtained as follows ($j=NC,\, CC$):
\begin{eqnarray}\label{eq:hadsigma}
d\sigma^\nu_{j}(E_\nu) & = & 
\sum_{i} \int dx \; q_i(x,Q^2) \; (d\hat \sigma_{0,(j)}^\nu+d\hat
\sigma_{v+s}^{j}) \nonumber \\
&+ & \sum_{i} \int_x^{1-\delta_s} \frac{d z}{z} \, q_i\left(\frac{x}{z},Q^2\right) \,
d\hat\sigma^{j}_c \nonumber \\
&+& \sum_i \int dx \; q_i(x,Q^2) \; d\hat \sigma^{j}_h \; ,
\end{eqnarray}
where the parton level cross section consists of the tree-level cross section, virtual, soft and
collinear ${\cal O}(\alpha)$ contributions (including the PDF counterterms)
and the real hard photon radiation contribution.  

\subsection{Fermion-mass effects}
\begin{figure}
\begin{center}
\includegraphics[width=0.4 \textwidth]{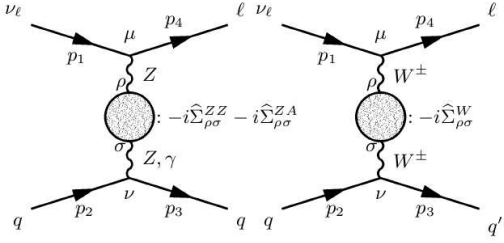}
\end{center}
\caption{Feynman diagrams for self-energy corrections to the NC and CC
  $\nu N$ production processes.}\label{SEC}
\end{figure}

We performed the calculation with and without including fermion-mass
effects and are considering the following two cases:
\begin{itemize}
\item[case 1:] All external fermions are considered to be massless and we only
  keep non-zero fermion masses as regulators of the collinear
  singularities.
\item[case 2:] The full muon and charm-quark mass dependence is taken into
  account, but light external fermions are treated as in the first case.
\end{itemize}
Fermion-mass effects in EW radiative corrections may not be numerically
negligible in this process, since the relevant parton-level energy scale
($q^2=\hat t$) and the fermion masses can be of the same order of
magnitude. This is illustrated 
below for the example of the $W$ self-energy correction ($\hat
\Sigma^W_{\rho\sigma}$) to the CC process shown in Fig.~\ref{SEC}. Its
contribution to the one-loop corrected matrix element at $\mathcal{O}(\alpha)$
reads
\begin{align}
  &\mathcal{M}_{virt}^{CC}=\nonumber \\
  & \,\, \frac{-e^2}{8s_W^2}\frac{\hs^W_{\rho \sigma}}{\left(q^2-M_W^2\right)^2}[\overline{u}_4 \gamma^\rho
  (1-\gamma_5) u_1][\overline{u}_3 \gamma^\sigma (1-\gamma_5) u_2].
\end{align}
With the renormalized $W$ self energy $\hs^W_{\rho \sigma}$ being decomposed in
transverse and longitudinal parts, $\hs^W_{\rho
  \sigma}=\left(g_{\rho\sigma}-\tfrac{q_\rho q_\sigma}{q^2}\right)
\hs^W_T+\tfrac{q_\rho q_\sigma}{q^2} \hs^W_L$, one finds the following
contribution to the NLO matrix element squared (up to terms of ${\cal O}(m_f^2/M_W^2)$):
\begin{align}
  2&{\rm Re}\mathcal{M}_{LO}^{CC*} \mathcal{M}_{virt}^{CC}=\frac{-4e^4}{s_W^4
    \left(\hat t-M_W^2\right)^3} \Bigg[p_1\idot p_2 \, p_3 \idot p_4 {\rm Re}\hs^W_T \nonumber \\
  &+\frac{m^2_4\left(m_2^2 p_1\idot p_3 - m_3^2 p_1\idot p_2 \right){\rm
      Re}\!\left(\hs^W_L-\hs^W_T\right)}{4\hat t}
  \Bigg]. \label{sew1}
\end{align}
If we consider massless fermions for the external legs (case 1), the
second term in Eq.~\eqref{sew1} vanishes. However, for massive fermions
(case 2) the longitudinal two--point function contributes to the physical
cross section. Since we work in the Feynman-'t Hooft gauge, we also had
to include the contributions from the would-be Goldstone bosons, which
are not explicitly shown here.  In the $s$-channel $W$ production
process such as gauge-boson production in Drell-Yan processes at the
Tevatron and the LHC, $\hat t$ is replaced with $\hat s$, so that the
second term is usually negligible. In $t$-channel deep inelastic
scattering, however, fermion-mass effects deserve a closer
investigation, especially in the small $\hat t$ region, which corresponds to
a small momentum fraction $x$. Note that similar effects also arise from
vertex and box corrections.

\begin{figure}
\begin{center}
\includegraphics[width=0.4\textwidth]{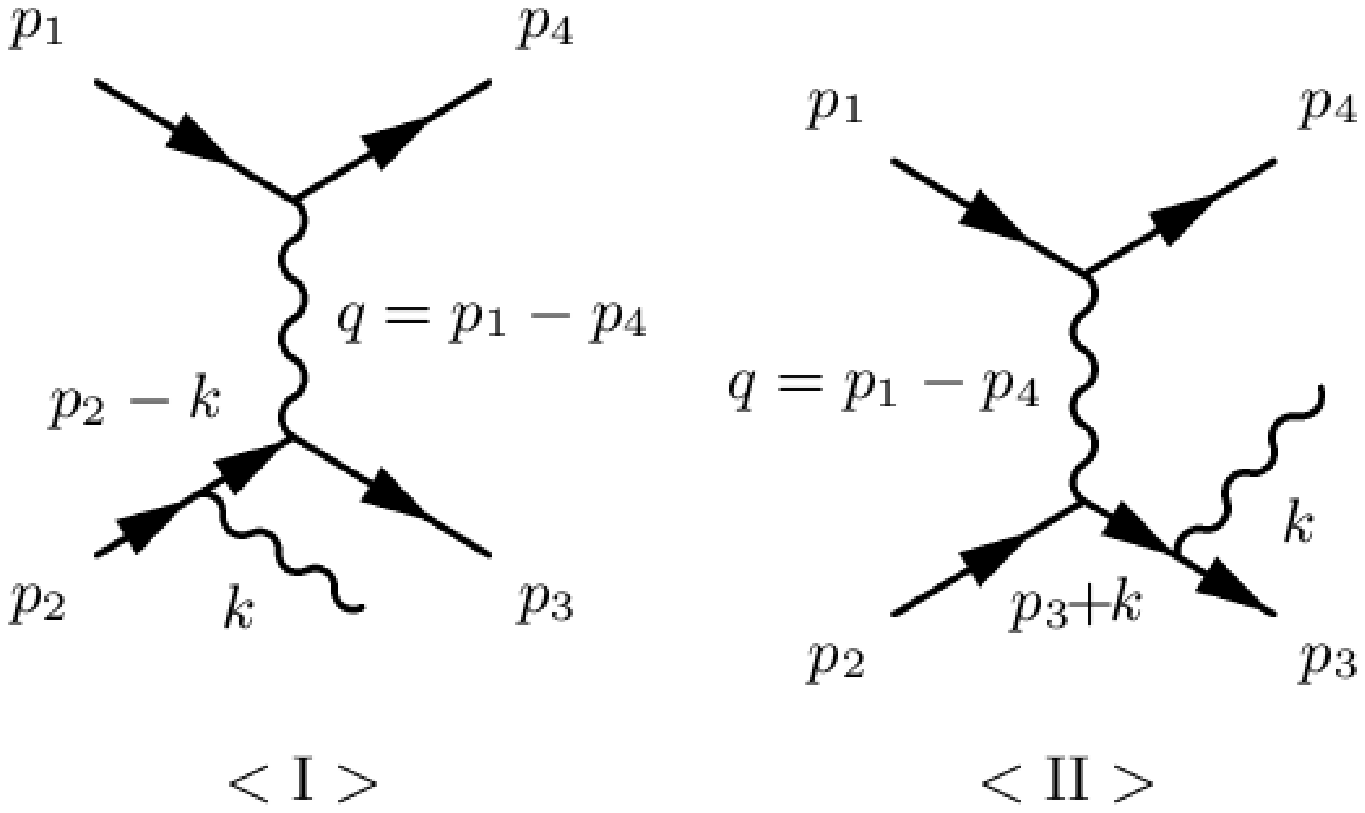}
\includegraphics[width=0.4\textwidth]{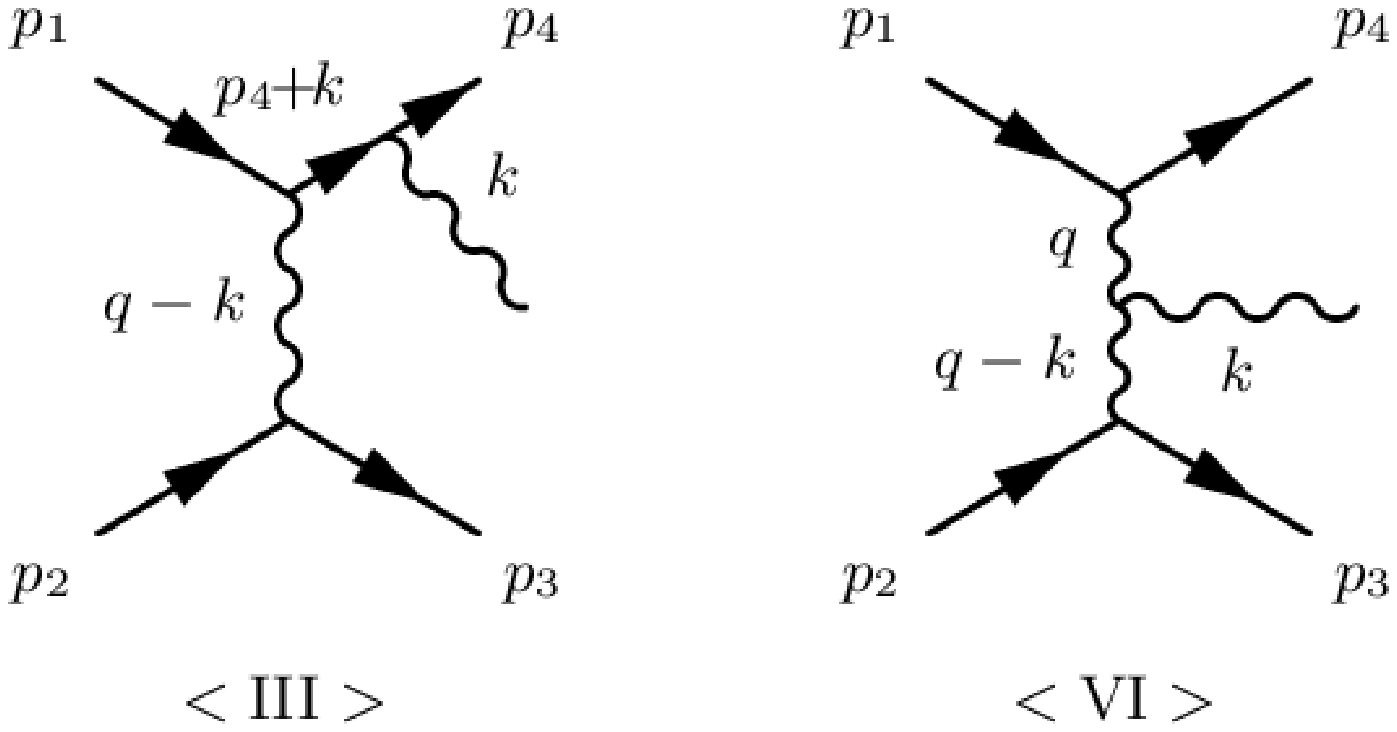}
\caption{Feynman diagrams for real photon radiation in the CC $\nu N$
  scattering process. Shown here are only $W^\pm$ exchange diagrams.}\label{real}
\end{center}
\end{figure}

\begin{figure}
\begin{center}
\includegraphics[width=0.4\textwidth]{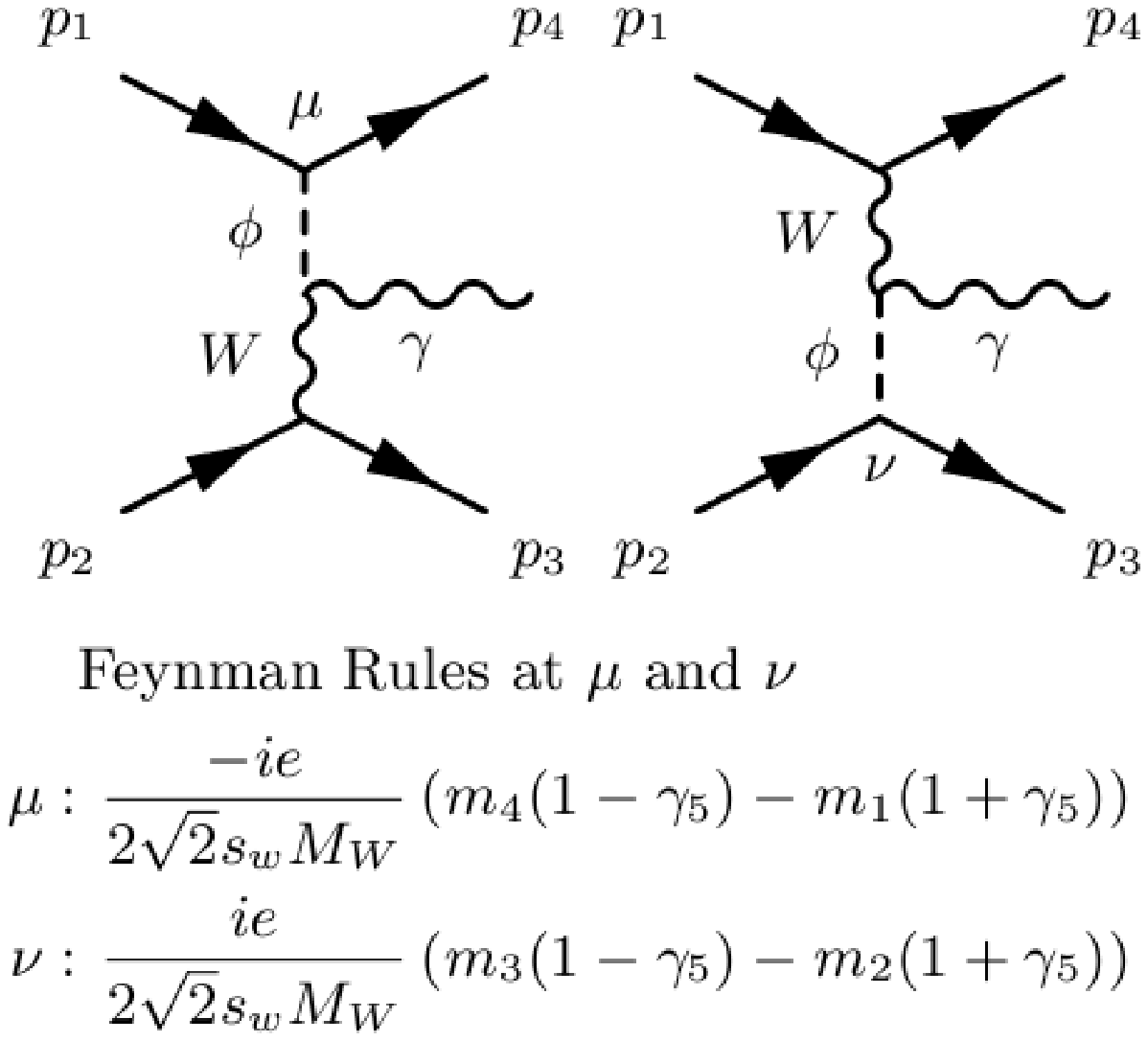}
\caption{Feynman diagrams for the mixed $W^\pm-\phi^\pm$ exchange contribution to real photon radiation in the CC $\nu N$
  scattering process, and the Feynman rules for the $\phi^\pm ff'$ coupling.}\label{sb}
\end{center}
\end{figure}

In case of real photon radiation amplitude-level fermion-mass effects only
arise in the CC $\nu N$ scattering process (the matrix elements to real photon
radiation in the NC process of cases 1 and 2 are identical).  The Feynman
diagrams for the $W^\pm$ exchange contribution are shown in Fig.~\ref{real},
and the corresponding matrix element (${\cal M}_r^{CC}$) can be written in
terms of $U(1)$-conserved leptonic and hadronic currents as follows
($\epsilon_{\rho}(k)$ denotes the photon polarization vector):
\begin{equation}
{\cal M}_r^{CC}= \frac{e^3}{8 s_w^2} \, \left[\frac{{\cal M}_{hadr}^{CC,\rho}}{(\hat
    t-M_W^2)}+\frac{{\cal M}_{lept}^{CC,\rho}}{(\hat t-M_W^2-2 k\idot
    q)}\right] \, \epsilon_{\rho}^*(k)
\end{equation}
with \\
\begin{align*}
&\mathcal{M}_{hadr.}^{\rm CC,\rho} \!=\! [\overline{u}_4  \gamma_\mu
\gamma_L u_1] [\overline{u}_3 \!\Big( \! \Gamma_\ra^{\mu\rho}\!+\!\Gamma_\rb^{\mu\rho}\!+\!\Gamma_\rda^{\mu\rho} \! \Big) \!\gamma_L
u_2]\! +\! \cancel{\mathcal{J}^\rho_m}
\end{align*}
\begin{align*}
&\mathcal{M}_{lept.}^{\rm CC,\rho} = [\overline{u}_4   \!\Big( \! \Gamma_\rc^{\mu\rho}+\Gamma_\rdb^{\mu\rho} \! \Big)
\gamma_L u_1] [\overline{u}_3 \gamma_\mu  \!\gamma_L u_2] -\cancel{\mathcal{J}^\rho_m}
\end{align*}
where $\gamma_L=1-\gamma_5$ and,
\begin{align*}
\Gamma_\ra^{\mu\rho}&=Q_2 \gamma^\mu \frac{p_2^\rho -\kslash \gamma^\rho/2}{-k\idot p_2}, \quad \Gamma_\rb^{\mu \rho}=Q_3 \frac{p_3^\rho+ \gamma^\rho \kslash/2}{k\idot p_3}\gamma^\mu\\
\Gamma_\rc^{\mu\rho}&=Q_4\frac{p_4^\rho+\gamma^\rho \kslash/2}{k\idot p_4} \gamma^\mu,\quad \Gamma_\rda^{\mu\rho}=\frac{\gamma^\mu q^\rho+\gamma^\rho k^\mu -g^{\mu \rho} \kslash}{-k\idot q}\\
\Gamma_\rdb^{\mu\rho}&=\frac{\gamma^\mu q^\rho-\gamma^\rho k^\mu+ g^{\mu \rho}\kslash}{k\idot q}, \qquad q^\mu=p_1^\mu-p_4^\mu 
\end{align*}
\begin{align*}
\mathcal{J}_m^\rho &=\frac{1}{2k\idot q}    \Big(
          m_1[\overline{u}_4 (1+\gamma_5) u_1][\overline{u}_3 \gamma^\rho (1-\gamma_5)  u_2] \\
& \qquad -m_4[\overline{u}_4 (1-\gamma_5) u_1][\overline{u}_3 \gamma^\rho (1-\gamma_5)  u_2] \\
& \qquad -m_2[\overline{u}_4  \gamma^\rho (1-\gamma_5) u_1][\overline{u}_3 (1+\gamma_5)  u_2] \\
& \qquad +m_3[\overline{u}_4  \gamma^\rho (1-\gamma_5) u_1][\overline{u}_3 (1-\gamma_5)  u_2]   \Big).
\end{align*}
where the subscripts, $\ra,\rb,\rc$,and $\rd$ correspond to the diagrams shown
in Fig.~\ref{real}. The fermion-mass dependence of ${\cal M}_r^{CC}$ described
by $\mathcal{J}^\rho_m$ vanishes when the contribution of the mixed
$\phi^\pm-W^\pm$ exchange diagrams shown in Fig.~\ref{sb} is included.  The
only surviving fermion-mass dependence at the amplitude-level is due to the
$\phi^\pm$ exchange diagrams which are, however, suppressed by ${\cal
  O}(m_f^2/M_W^2)$.

\subsection{Treatment of numerical instabilities}

\begin{figure}
\includegraphics[width=0.38\textwidth]{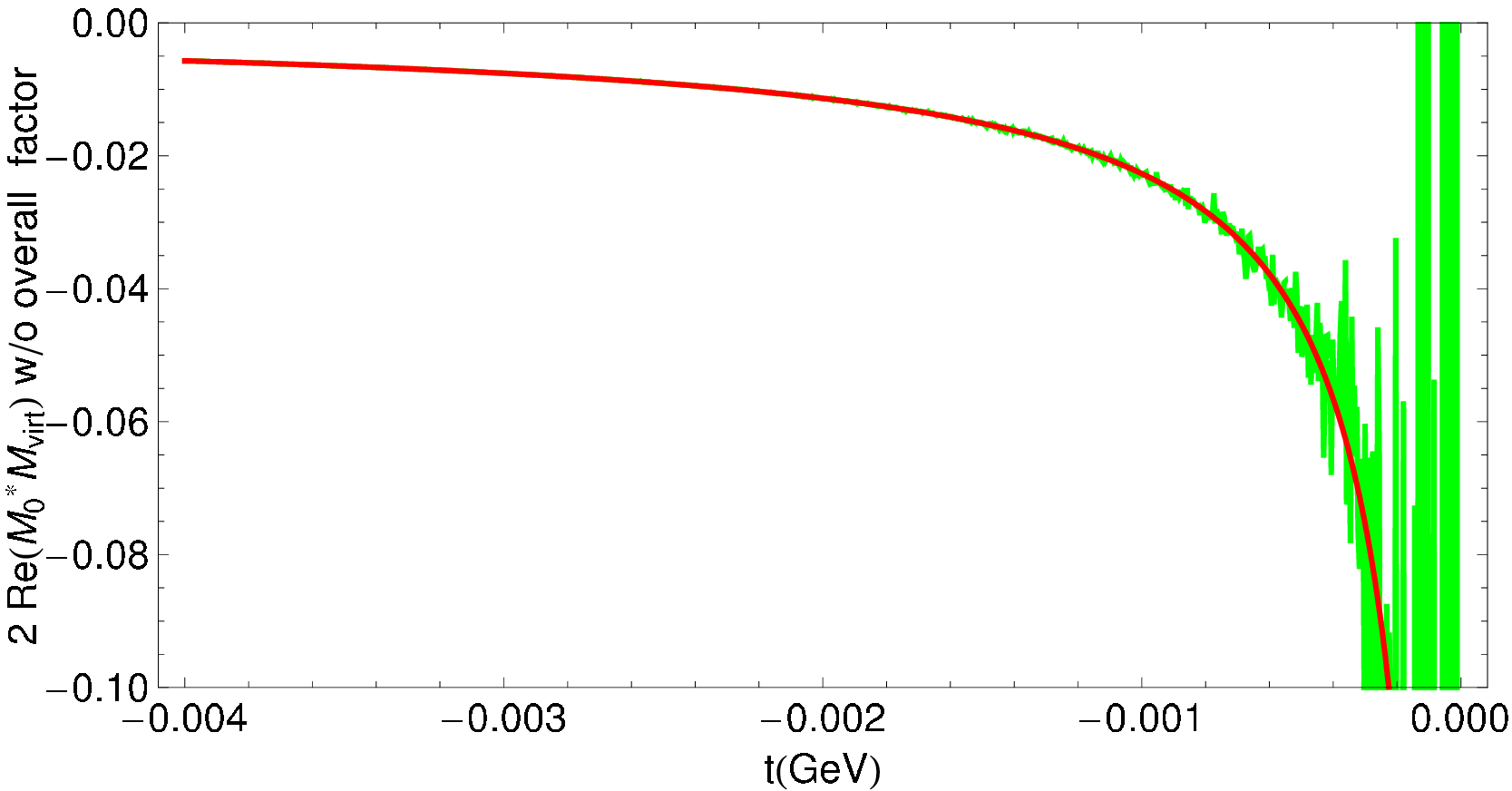}\includegraphics[width=0.13\textwidth]{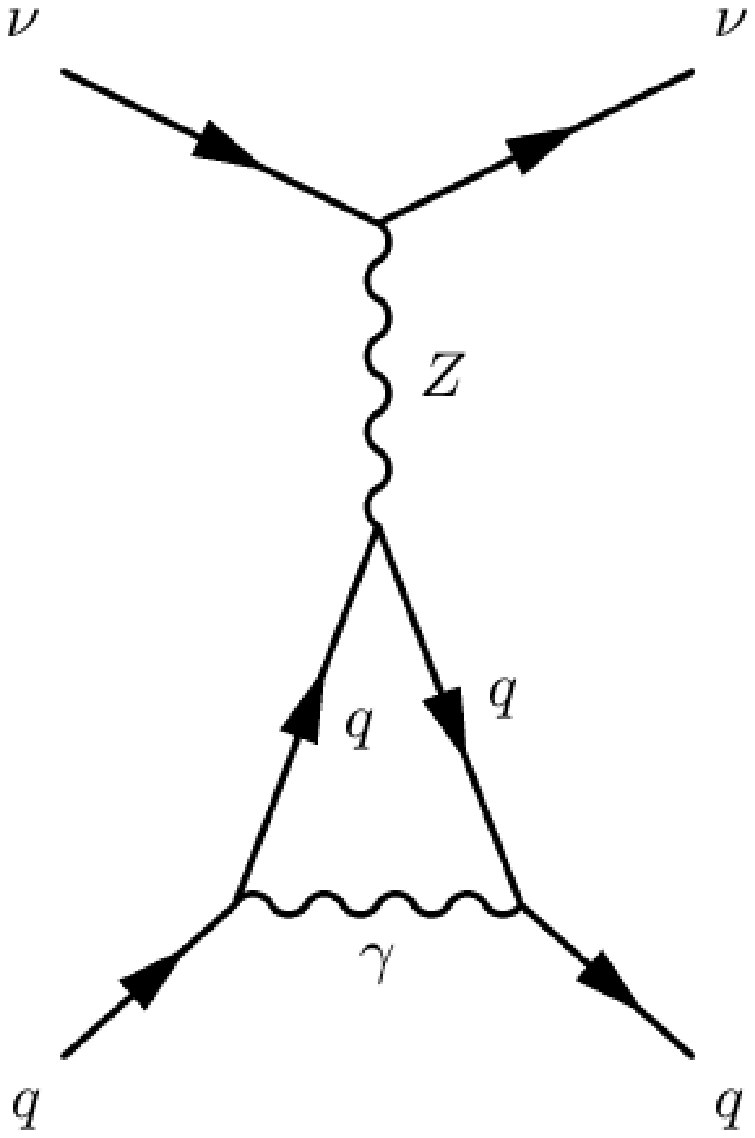}
\caption{Numerical instabilities occurr in fermion-fermion-photon exchange
  vertex corrections (green curve) owing to the photon propagator, 
while the Taylor expansion yields numerical stable results (red curve). }\label{ni_exp}
\end{figure}

It is well-known (see, e.~g., Ref.~\cite{Diener:2003ss}) that the EW NLO cross section to
the NC process suffers from a numerical instability at small values of $\hat t$
owing to photon exchange diagrams such as shown in Fig.~\ref{SEC} and Fig.~\ref{ni_exp}. As a
remedy of this kind of instability, we apply a Taylor expansion around
small $\hat t$. In Fig.~\ref{ni_exp} we illustrate the stability of this
expansion.

\begin{figure}
\includegraphics[width=0.4\textwidth]{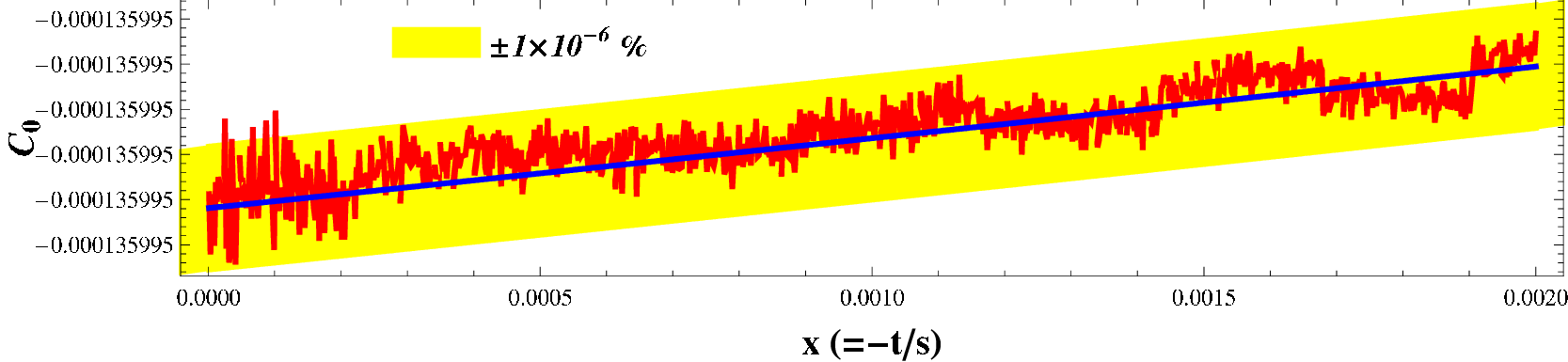}
\includegraphics[width=0.4\textwidth]{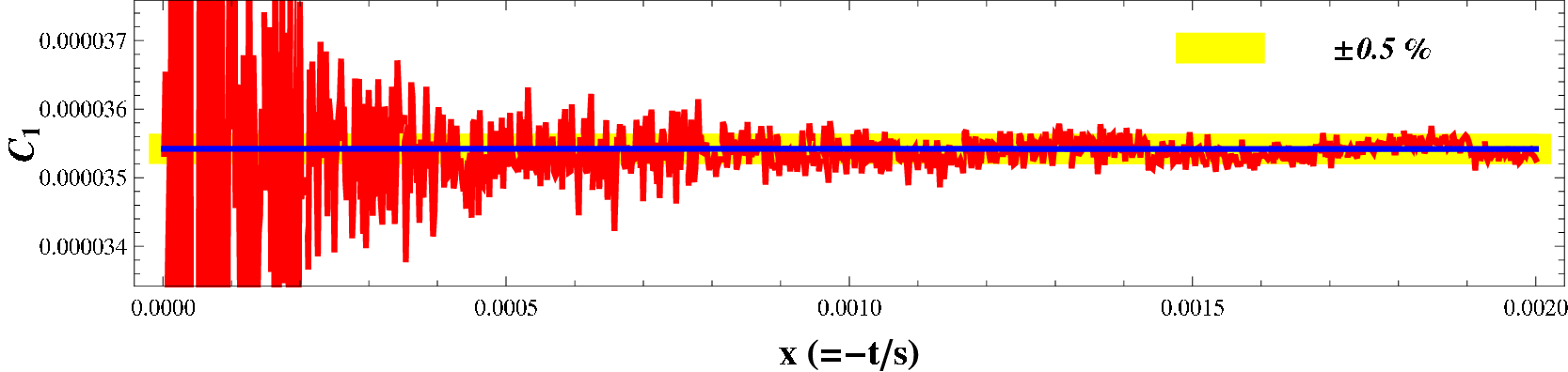}
\includegraphics[width=0.4\textwidth]{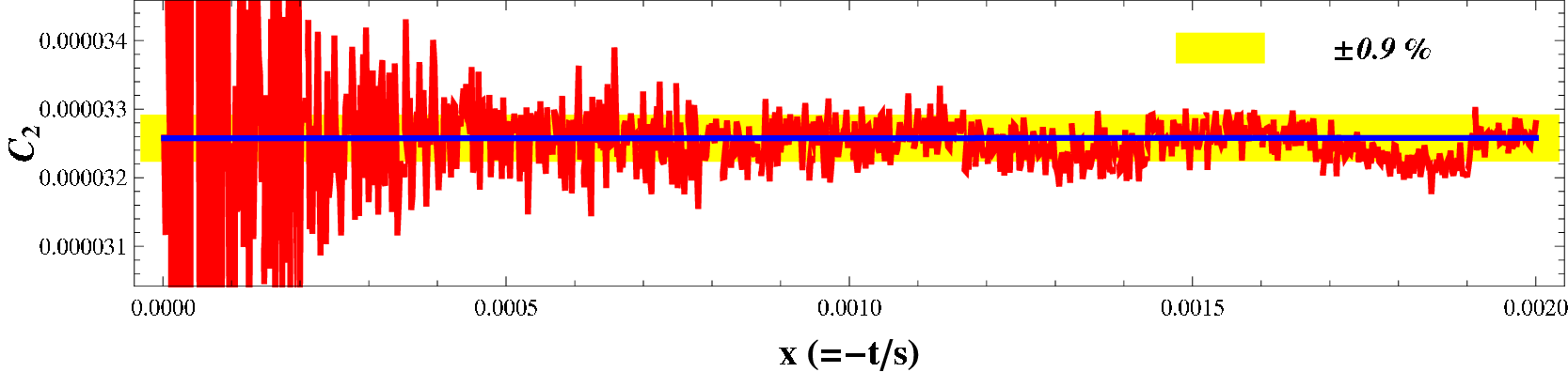}
\includegraphics[width=0.4\textwidth]{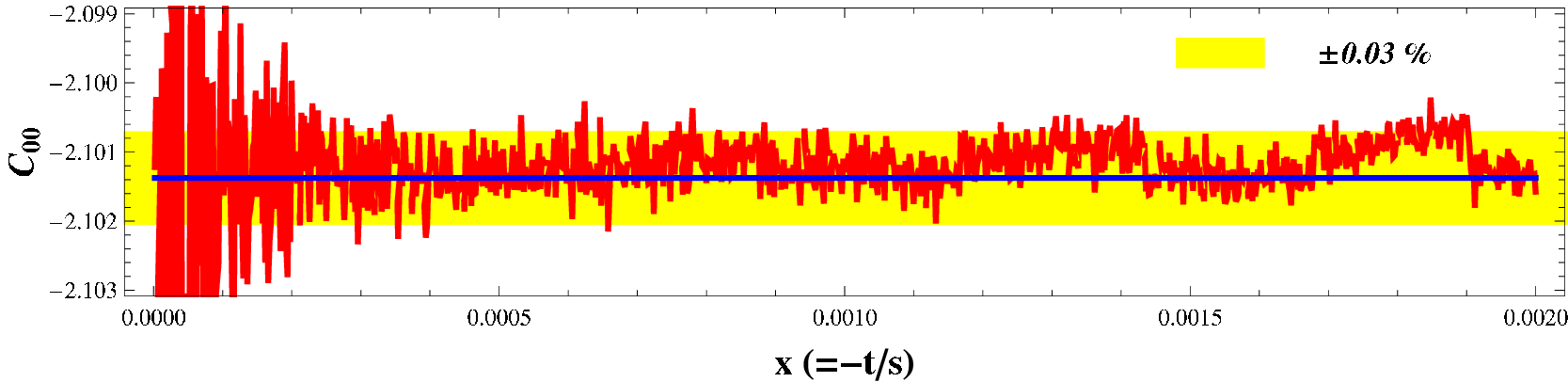}
\includegraphics[width=0.4\textwidth]{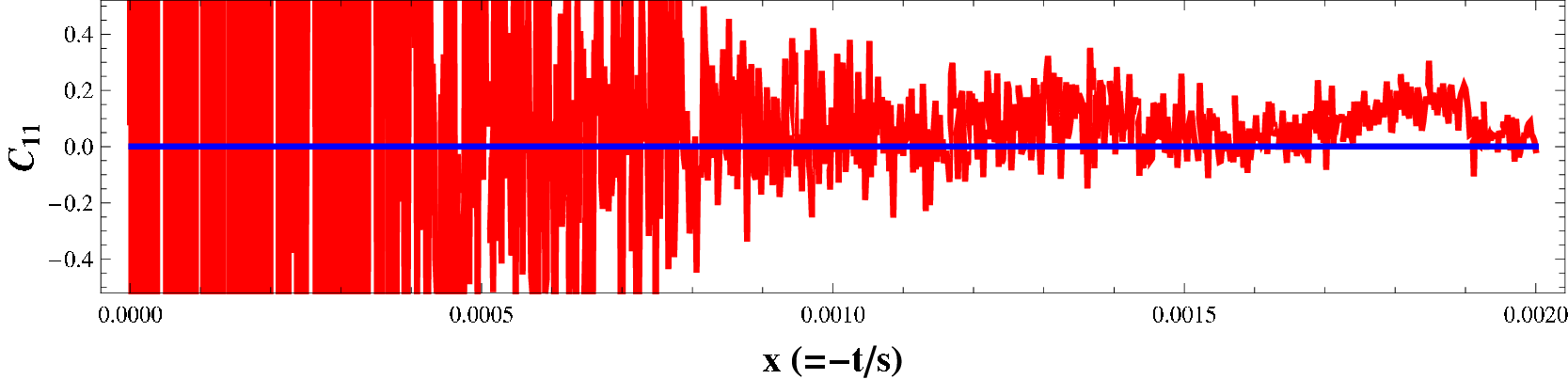}
\includegraphics[width=0.4\textwidth]{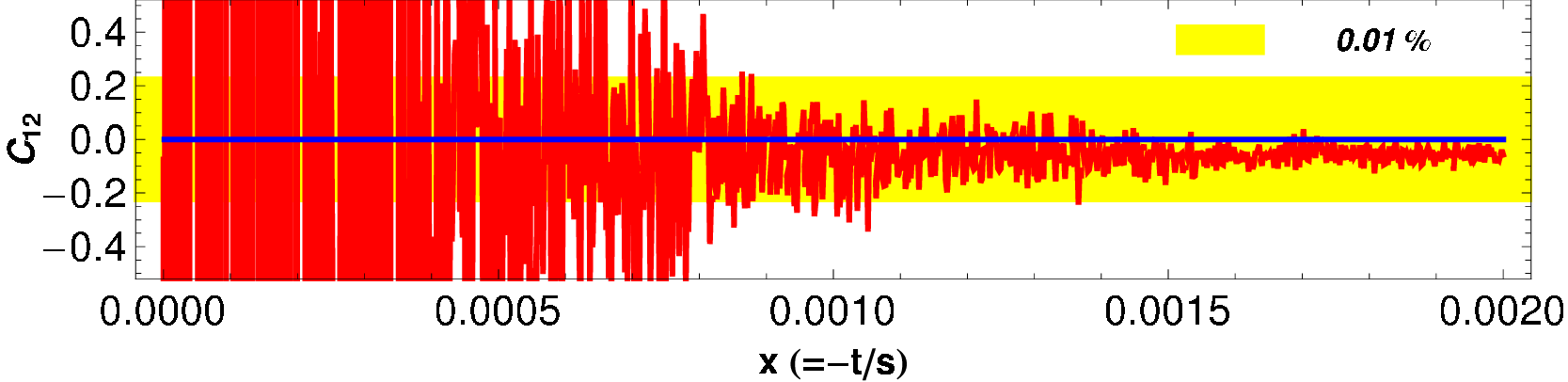}
\includegraphics[width=0.4\textwidth]{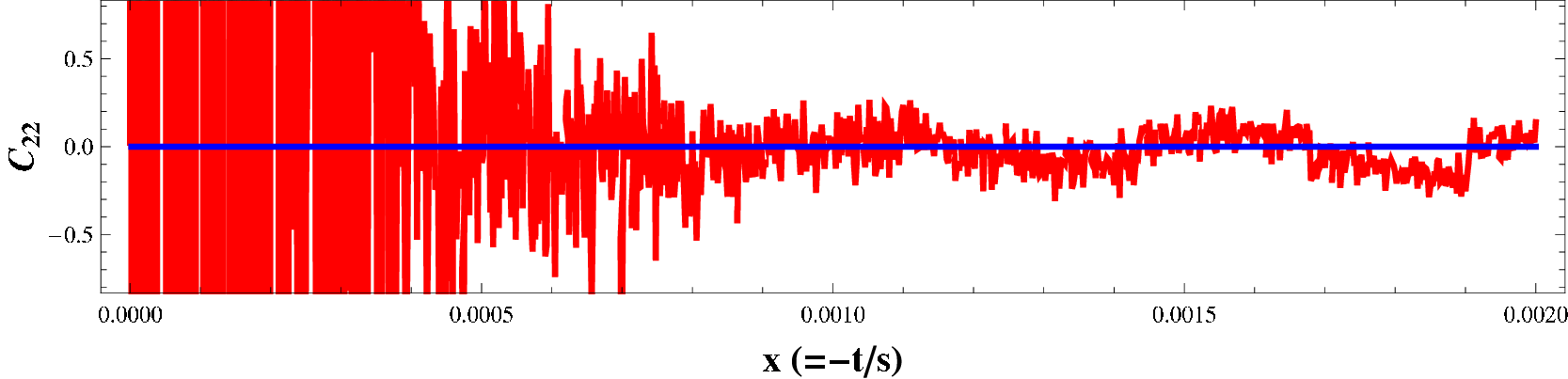}
\caption{The standard Passarino-Veltman
  reduction exhibits a numerical instability (red curve) while
  Eq.~\eqref{pr} provides stable results (blue curve). Shown are the
three point functions as a function of the variable $x=-t/s$. }\label{figC}
\end{figure}

Box corrections exhibit numerical instabilities originating from vanishing Gram
determinants at small kinematic variables when the standard Passarino-Veltman
reduction formalism~\cite{Passarino:1978jh} is employed to determine
the coefficient functions of vector and tensor four-point
integrals. Especially, the crossed box contribution in the low
$x$-region suffers from these instabilities which can be traced back to
numerical unstable coefficients of three-point integrals. In the following we present
a reduction formalism which yields stable results as illustrated in
Fig.~\ref{figC} for case 1 (see also, e.~g., \cite{Denner:2005nn} for an alternate solution).  In
phase-space regions of small kinematic variables, the coefficients of the
three-point vector and tensor integral 
\begin{align*}
&C^\mu,C^{\mu \nu }=\\
&\frac{(2 \pi \mu)^{4\!-\!D}}{i\pi^2} \!\int \!\!d^D \!q \frac{q^\mu,q^\mu q^\nu}{[q^2\!-\!m_0^2][(q\!+\!p_1)^2\!-\!m_1^2][(q\!+\!p_2)^2\!-\!m_2^2]},
\end{align*}
defined as
\begin{align*}
C^\mu&=p_1^\mu C_1 + p_2^\mu C_2, \\
C^{\mu \nu}&=g^{\mu\nu}C_{00}+p_1^\mu p_1^\nu C_{11}+p_1^\mu p_2^\nu C_{12}+p_2^\mu p_1^\nu
C_{21}+ \\
&\phantom{\,=}  p_2^\mu p_2^\nu C_{22},
\end{align*}
can be approximated in terms of two-point functions as follows
\begin{align*}
C_1&= \alpha B^1_1 +2 \alpha^2 \left[  B_{00}^1-B_{00}^2+(p_1^2-p_1\.p_2)B_{11}^1 \right] \nonumber \\
&\qquad+\mathcal{O}\left((p_1^2)^2,(p_2^2)^2,(p_1\.p_2)^2\right) \nonumber \\
C_2&= -\alpha B^2_1 +2 \alpha^2 \left[ B_{00}^2-B_{00}^1+(p_2^2-p_1\.p_2)B_{11}^2 \right] \nonumber \\
&\qquad+\mathcal{O}\left((p_1^2)^2,(p_2^2)^2,(p_1\.p_2)^2\right)
\end{align*}
\begin{align*}
C_{00}&= \alpha ( B_{00}^1 - B_{00}^2 )+2 \alpha^2 \Big[ (p_1^2-p_1\.p_2)B_{001}^1 \nonumber \\
&\qquad + (p_2^2-p_1\.p_2)B_{001}^1\Big] +\mathcal{O}\left((p_1^2)^2,(p_2^2)^2,(p_1\.p_2)^2\right) \nonumber \\
C_{11}&= \alpha (B_{11}^1) +2 \alpha^2 \left[ 2B_{001}^1+(p_1^2-p_1\.p_2) B_{111}^1 \right]  \nonumber \\
&\qquad+\mathcal{O}\left((p_1^2)^2,(p_2^2)^2,(p_1\.p_2)^2\right)\nonumber \\
C_{12}&=C_{21}= -2 \alpha^2 \left[ B_{001}^1+B_{001}^2\right]  \nonumber \\
&\qquad+\mathcal{O}\left((p_1^2)^2,(p_2^2)^2,(p_1\.p_2)^2\right) \nonumber \\
C_{22 }&= -\alpha (B_{11}^2) +2 \alpha^2 \left[ 2B_{001}^2+(p_2^2-p_1\.p_2) B_{111}^2\right] \nonumber \\
&\qquad+\mathcal{O}\left((p_1^2)^2,(p_2^2)^2,(p_1\.p_2)^2\right)
\end{align*}
and the scalar three-point integral reads:
\begin{align}\label{pr}
C_0&=\alpha (B^1_0\!-\!B^2_0) \!+\!2 \alpha^2 \left[ (p_1^2\!-\!p_1\.p_2) B_1^1\!+\!(p_2^2\!-\!p_1\.p_2) B_1^2\right]\nonumber \\ &\qquad+\mathcal{O}\left((p_1^2)^2,(p_2^2)^2,(p_1\.p_2)^2\right).
\end{align}
Here,
\begin{align*}
\alpha=\frac{1}{m_1^2-m_2^2-p_1^2+p_2^2}, \quad
B^i_{\mu \nu \cdots}=B_{\mu \nu \cdots}(p_i^2,m_0^2,m_i^2).
\end{align*}
The detailed derivation of these expressions can be found
in~\cite{park}. In Fig.~\ref{figC} we show a comparison of these two
derivations of the $C_{ij}$ functions in the critical phase space
region, i.~e. at small $x$.

\section{Numerical results}

The measured value of $\sin^2 \theta_W$ can be extracted from the
Paschos-Wolfenstein relation~\cite{Paschos:1972kj}
\begin{align}
R&=\frac{\sigma^\nu_{NC}(\nu N \rightarrow \nu X)-\sigma^{\bar \nu}_{NC}(\bar{\nu} N \rightarrow \bar{\nu} X)}
{\sigma^{\nu}_{CC}(\nu N \rightarrow \ell X)-\sigma^{\bar{\nu}}_{CC}(\bar{\nu} N \rightarrow \bar{\ell} X)} \nonumber \\
&=\rho^2 \left(\frac{1}{2} - \sin^2\theta_W \right)
\end{align}
In the on-mass-shell renormalization scheme, $\sin^2 \theta_W$ is
related to the masses of both $W$ and $Z$ bosons as follows:
\begin{align}
\sin^2\theta_W = 1 -\frac{M_W^2}{M_Z^2}.
\end{align}
In order to determine the impact of the higher-order corrections under
investigation on the extracted value of $\sin^2 \theta_W$, we use the
following expression~\cite{Bardin:1986bc}:
\begin{equation}
\Delta \sin^2 \theta_W =\frac{\tfrac{1}{2}-\sin^2 \theta_W+\tfrac{20}{27} \sin^4 \theta_W}
{1-\tfrac{40}{27} \sin^2 \theta_W} \left(\delta R^\nu_{NC}+\delta R^\nu_{CC}\right),
\end{equation}
where the ratios of $\delta R^\nu_{NC}$, $\delta R^\nu_{CC}$ and $ R^\nu_0$ are defined as
\begin{align*}
R_0^\nu=\frac{\sigma^\nu_{0,NC}}{\sigma^\nu_{0,CC}}, \quad \delta R_{NC}^\nu=\frac{\delta \sigma^\nu_{NC}}{\sigma^\nu_{NC}}
, \quad \delta R_{CC}^\nu=-\frac{\delta \sigma^\nu_{CC}}{\sigma^\nu_{CC}}.
\end{align*}

For the numerical evaluation we use the same input parameters and PDF's
as Ref.~\cite{Diener:2003ss}.

\begin{figure}
\begin{center}
\includegraphics[width=0.45\textwidth]{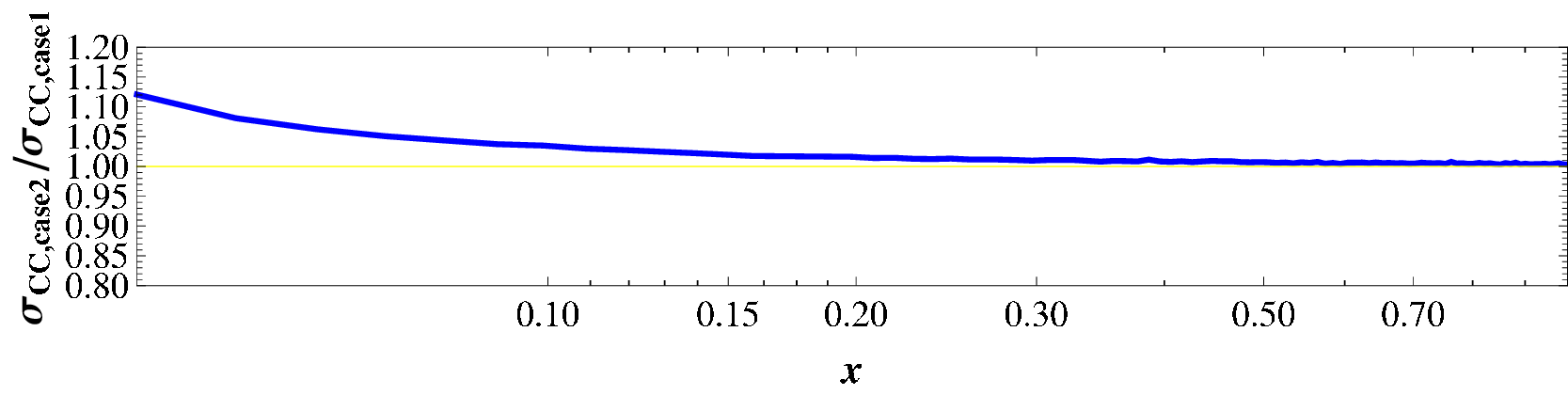}
\includegraphics[width=0.45\textwidth]{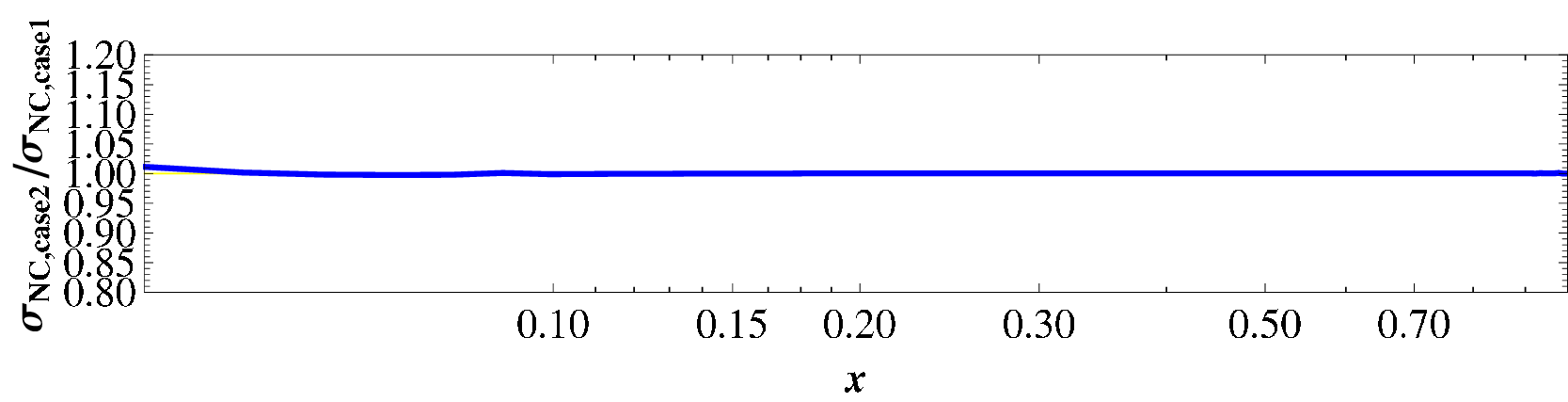}
  \caption{Ratio of $\nu N$ scattering cross sections calculated with
    massless and massive external fermions for both the CC and NC
    production process.}\label{figR}
\end{center}
\end{figure}

As discussed earlier, for the massless calculation (case 1), we neglect
fermion masses whenever it is possible, i.~e. we take their masses only
for regularizing singularities, while, in the massive case (case 2), we
keep the muon and charm-quark masses. Note that following results are
\emph{preliminary}. As can be seen in Fig.~\ref{figR}, where we compare
the $\nu N$ scattering cross sections calculated in both cases for both
CC and NC scattering processes, fermion-mass effects are only visible at
small $x$ and are more pronounced in the CC case.  How this may
translate into the extracted value of $\sin^2\theta_W$ is illustrated in
Tab.~\ref{res}. More detailed studies are under way.

\begin{table}[h]
\begin{center}
\caption{Preliminary results obtained with the {\it MRST2004QED} PDF set
  and a cut on $y=- \hat t/\hat s \ge 0.12$.}
\begin{tabular}{c|c|c|c|c}
\hline
                & $R_o^\nu$ & $\delta R_{NC}^\nu$ & $\delta R_{CC}^\nu$ & $\Delta \sin^2\theta_W$  \\
\hline Case 1 & 0.30638   & 0.0527              & -0.0916             & -0.0182 \\
\hline Case 2  & 0.31477   & 0.0548              & -0.1059             & -0.0258 \\
\hline
\end{tabular}
\label{res}
\end{center}
\end{table}

\section{Conclusion}

Deep-inelastic neutrino--nucleon scattering provides an excellent
testing ground for the electroweak SM, complementary to $e^+e^-$ and
hadronic colliders.  Measurements of electroweak parameters in
neutrino--nucleon scattering are not only comparable in precision but
they also probe the EW SM at many orders of magnitude in $\hat t=q^2$,
i.e. the parton-level momentum transfer in these processes.

The NuTeV collaboration used the calculation of Ref.~\cite{Bardin:1986bc},
which is based on a massless fermion approximation, and did not include
the entire set of electroweak $\mathcal{O}(\alpha)$ corrections.
Seventeen years later, a complete calculation of the
$\mathcal{O}(\alpha)$ corrections to neutrino-nucleon scattering became
available \cite{Diener:2003ss}. In a follow-up paper
\cite{Diener:2005me}, leading higher order corrections, i.e. beyond
one-loop, have been included as well. In Ref.~\cite{Diener:2003ss}, the
discussion focused on the EW input scheme dependence of $\sin^2
\theta_W$ measured in neutrino--nucleon scattering. They concluded that
the theoretical uncertainty due to missing higher-order corrections has
been underestimated by the NuTeV collaboration, and, thus, is a
potential source for at least part of the observed discrepancy.

In this study we focus on another potential source of a theoretical
uncertainty, which has not been considered before. i.e.  the effects of muon
and charm-quark masses in the calculation of electroweak corrections. We
calculated the complete electroweak $\mathcal{O}(\alpha)$ corrections to
neutrino--nucleon scattering with and without taking into account these
fermion-mass effects. We studied their impact on $\sin^2 \theta_W$ as
extracted from the $\nu N$ scattering cross section by the NuTeV
collaboration.  We found non-negligible differences in $\sin^2\theta_W$ when
using our calculation with and without considering non-zero muon and
charm-quark masses.  However, a more realistic study is needed including a
simulation of the detector resolution, for instance, to determine whether
these effects can account for part of the NuTeV anomaly. Such a study is
currently in progress.


\begin{acknowledgments}
  We are grateful to Kevin McFarland for fruitful discussions and
  guidance concerning experimental issues.
  The work of K.~P. is supported by the U.S. Department of Energy under
  grant DE-FG02-04ER41299. This research is also supported by the
  National Science Foundation under grants No.~NSF-PHY-0547564 and
  NSF-PHY-0757691. The work of D.~W. is presently also supported by a DFG
  Mercator Visiting Professorship.
\end{acknowledgments}

\bigskip 

\end{document}